\documentclass[12pt]{article}
\begin{document}
\sffamily
\begin{center}
{\large\bf Supersolid and the non-uniform condensate}\\~\\

N. Kumar\\
Raman Research Institute, Bangalore 560080, India

\end{center}
\vspace{0.50cm}

\begin{abstract}
We construct a model of non-uniform condensate having a spatially modulated complex order parameter that makes it kinematically an x-ray solid, i.e., a real mass density wave, but one admitting an associated superfluid flow. Intrinsic to this state is a non-classical translational inertia which we derive for the case of a potential flow. Connection to the non-classical rotational inertia observed in recent experiments on solid helium-4 is discussed. Our semi-phenomenological treatment suggests a flow-induced supersolid-to-superfluid transition. 
\end{abstract}
\baselineskip=24pt
Recent experiments of Kim and Chan [1,2] on solid helium-4 at very low temperatures have strikingly revealed a non-classical rotational inertia that seems intrinsic to it, much as it is in the case for superfluid He II. Such a supersolid was indeed predicted much earlier on theoretical grounds as a plausible concomitant of a quantum crystal with delocalized defects, or of a Bose-Einstein condensate [3-5], and had motivated years of research [6]. The non-classical inertial effect was, however, estimated to be very small, and direct tests were therefore suggested [5]. Thus, the question {\em ``can a solid be superfluid?"}, raised some 35 years ago [5] has now been finally answered in the affirmative. In this work we explicitly construct a complex order-parameter that underlies a real mass density wave modeling the supersolid, and demonstrate analytically its necessarily non-classical inertia $-$ the signature of a supersolid. We also predict a flow-induced supersolid-to-superfluid transition.

In order to motivate an order-parameter approach to the supersolid, let us first recall that geometrically a solid is a periodic spatial modulation of matter density. Such a crystalline solid structure will be revealed kinematically in its characteristic x-ray diffraction peaks $-$ we may call it an x-ray solid [3]. This kinematic description, however must be supplemented by the energetics of its stability against deformation. We will show below that both these conditions are  realized in our order-parameter description of the supersolid admitting a non-zero superfluid flow relative to the laboratory frame in which the density wave is at rest. We will specifically consider a potential flow and show that the translational inertia of the supersolid is smaller than its literal mass. The potential flow is consistent with the geometry (topology) of the experiments [1,2] in which the flow is confined to the narrow annular space between two co-axial cylinders, and has quantized Onsager-Feynman circulation [= $nh/m$] taken around the annulus. The situation can be more complicated for a simply connected topology with quantized vortices in it. We will then comment on its relation to the diminished rotational inertia observed in the above cited experiments. 

Our construction of the model supersolid state for a system of interacting Bose particles such as $^4$He is informed by the following physical considerations. The interparticle interaction (repulsion) is known to deplete the condensate. The interaction, however, favours large fluctuations, i.e., an instability, towards long-range diagonal order corresponding to a crystalline solid, as evidenced by the peak in the liquid helium static structure factor at the roton minimum wavevector. Failure to form a solid (under its saturation vapour pressure) is, however, prevented by the large zero-point kinetic energy due to the high value of the de Boer quantum parameter (= $\hbar / \sigma (m\epsilon)^{1/2}$, where $\sigma$ and $\epsilon$ are, respectively, the range and the depth of the interaction potential). It solidifies only under pressure that offsets the zero-point pressure through the PV term in the Gibbs potential.  The large zero-point amplitude comparable to the interparticle separation, however, persists and the associated delocalization leads to the possibility of an off-diagonal long-range order co-extensive with the diagonal crystalline order, i.e., the supersolid phase.  With this picture in mind, consider a system of scalar Bose particles at zero temperature, and let its number density  $(n({\bf x}))$ be modulated in space as a real density wave 
$$
n({\bf x})  =  n_0 + \sum_{\bf g}\,\,n_{\bf g} \,\,\cos{\bf g \cdot x}, \eqno(1)		
$$
where summation over ${\bf g}$ spans the reciprocal lattice vectors for the periodically modulated condensate. 

The complex order parameter $\psi(x)$ underlying the above real density wave modulation, and obeying in general the Gross-Pitaevskii equation [7], is then 
$$
\psi({\bf x}) = \sqrt{n({\bf x})} exp(i\phi({\bf x})), \eqno(2)
$$
where $\psi({\bf x})$ is as usual the macroscopic wavefunction  obtained by Bose condensing a macroscopic number ($N$) of the Bose particles into a single one-particle state. Our simple model assumes complete condensation at zero temperature.
Now, associated with this complex order parameter $\psi({\bf x})$ there is a particle-number current density ${\bf j(x)}$, given by
$$
{\bf j(x)} = \big(\frac{-i\hbar}{2m}\big ) \big( \psi*({\bf x}) {\bf \nabla} \psi({\bf x}) - c.c.\big ) \equiv \frac{\hbar}{m} n({\bf x}) {\bf \nabla}\phi,  \eqno(3)$$
In the laboratory frame in which the real density modulation is assumed to be pinned still. Thus, we have the stationary flow with  ${\bf \nabla}~.~{\bf j}({\bf x}) = 0$, which determines ${\bf j}({\bf x})$ for the given boundary condition. The phase $\phi({\bf x})$ can then be obtained by solving 
$$
{\bf \nabla}\phi  =  \big(\frac{m}{\hbar}\big ) \frac{{\bf j}({\bf x})}{n_0 + \sum n_{\bf g} {\cos {\bf g}} . {\bf x}}~~~. \eqno(4)
$$

Our semi-phenomenological treatment of the supersolid phase, and of the possible supersolid-superfluid transition is now in principle as follows. With an appropriate choice for the dominant reciprocal lattice vectors for the density wave in Eq. (1), corresponding to the peak in the liquid-helium static structure factor [8], and for a given flow ${\bf j}({\bf x})$, we calculate the kinetic energy associated with the underlying complex order parameter $\psi({\bf x})$.  The use of a single dominant set of wavevectors, of course, means considerable delocalization of the atoms about the nominal lattice sites. This is, however, all the more justified for the solid helium than for a classical solid for reasons of large zero-point amplitude for the $^4$He atoms.  The part of the kinetic energy involving the flow ${\bf j}({\bf x})$ quadratically can now be expressed in terms of the total mechanical momentum associated with the flow. The real density modulation, of course, remains at rest in the laboratory frame. This at once identifies the inertia associated with the flow that will turn out to have a non-classical value. It will involve the modulation amplitude $\eta_g$ to be determined by the overall minimization of the Gibbs free energy. The part of the kinetic energy not involving the flow is to be identified as the zero-point energy due to the extent of localization implied by the density modulation. It will contribute an important part to the Gibbs free energy, and is known to prevent solidification of $^4$He under its saturation vapour pressure. It has to be off-set by an external pressure. Thus, we first consider the usual free energy where we follow the conventional mean-field theory of liquid-solid phase transition, but add to it the kinetic energy associated with the complex order parameter that underlies the real mass density wave as discussed above. Minimization with respect to $\eta_g$ then determines the density-wave amplitudes and the phase transition involved. The above procedure is straightforward in principle, but algebraically cumbersome in 3 dimensions. The ideas relevant to the supersolid phase are, however, contained essentially in the 1-dimensional model in the mean-field sense. This is the case we will now treat analytically. 

Specializing Eqs. (1-4) to the 1D case, and retaining the single dominant density wave for the wavevector of magnitude $g$ [8], we have from Eq. (4) for the phase $\phi({\bf x})$ [9] 
$$
\phi(x) = \big(\frac{2mj_0}{\hbar g n_0}\big ) \frac{1}{\sqrt{1-\eta_g^2}} \,\,\,arctg \big(\frac{{\sqrt{1 - \eta_g^2}} tg(gx/2)}{1 + \eta_g}\big ), \eqno(5)
$$
where we have set ${\bf j}({\bf x}) = j_0 = $ constant for the uniform stationary flow.

The kinetic energy $E_K$ associated with the complex order parameter $\psi(x)$ is readily calculated to be
$$
E_K = \frac{\hbar^2}{2m} \int_0^L \left | \frac{d\psi}{dx} \right |^2 dx 
$$
$$
= \frac{\pi^2}{2ma^2} \hbar^2 N(1 - \sqrt{1 - \eta_g^2}) + \frac{N j_0^2}{2n_0^2} m  \frac{1}{\sqrt{1 - \eta_g^2}} \equiv E_{\rm zero-point} + E_{\rm flow}, \,\, \eqno(6)$$
where $E_{\rm flow}$ is associated with the flow ($j_0$), and can be re-written as
$$
E_{\rm flow} = \frac{P^2}{2M\sqrt{1 - \eta_g^2}}, \eqno(7)
$$
with $P = L mj_0 = $ the total momentum associated with the flow, and $M = mn_0L = $ the total literal mass of the system. Thus, we can identify $M\sqrt{1 - \eta_a^2}$ with the associated inertial mass of the supersolid which is non-classical for $\eta_g \neq0$:
$$
M_{\rm non-classical} = M(1 - \eta_g^2)^{1/2} < M \eqno(8)
$$
which depends on the depth of modulation $\eta_g$ (to be determined from free-energy minimization).
Equation (8) shows that the non-classical inertia associated with the flow decreases monotonically with $\eta_g$ from $M$ (for $\eta_g = 0$, no solid-like density modulation) to zero (for $\eta_g = 1$, a solid). Such a trend is clearly  reasonable physically. 
The term $E_{\rm zero-point}$ in the expression for the kinetic energy $E_K$ in Eq. (6) is the zero-point energy due to the extent $(\sim 2\pi/g$) of localization associated with the spatial modulation at wavevector magnitude $g$.

Now, we turn to the free energy $F_0[\eta_g]$ (actually energy at zero temperature here with the entropy term omitted) whose minimization should give the density modulation $\eta_g$ and thus determine the phases and the phase transition. As discussed above, we have in the single dominant density-wave meanfield approximation [8]
$$
F_0/N = r_0 \eta_g^2 + u_{04} \eta_g^4 + u_{06} \eta_g^6 + \cdots , \eqno(8)
$$
where the subscript '0' denotes no flow ($j_0 = 0$)). (Note the omission of the cubic term on the R.H.S. of Eq. (8), well known in the context of classical liquid-solid transition in the 3D case where for a dominant $|{\bf g}|$ it is possible to choose three vectors forming an equilateral triangle for, e.g., an f.c.c.  lattice [8]. For a 1D case, however, this is not possible).  To this now we must add $E_K$ for a non-zero flow as obtained above. Also, a $PV$ term is to be added, where $P$ is the external pressure the system is subjected to.  This term can be readily shown to give $-\delta n_0 P$ in the Gibbs free energy for our 1D model, where $\delta$ is the reduction in the lattice spacing due to $P$. (For the 1D case here, $P$ is, of course, an external force). Thus, we have for the Gibbs free energy $G$ at zero temperature in the presence of flow and external pressure\\
$G \equiv F_0 + E_K + PV\\
~~~~ = r\eta_g^2 + u_4 \eta_g^4 + u_6 \eta_g^6 + \cdots\\
$
with \\
$r = r_0 + \frac{\pi^2\hbar^2}{4ma^2} + \frac{mj_0^2}{4n_0^2} - \delta n_0 P + \cdots $\\
$u_4 = u_{04} + \frac{1}{16}\,\,\frac{\pi^2\hbar^4}{ma^2} + \frac{3}{16}\,\,\frac{m}{n_0^2} j_0^2 + \cdots$\\
$u_6 = u_{06} + \frac{\pi^2\hbar^2}{32ma^2} + \frac{5mj_0^2}{32n_0^2} + \cdots, ~~~~~~~~~~~~~~~~~~~~~~~~~~~(9) $\\
where we have expanded $E_K$ also in powers of $\eta_g$ and collected the coefficients of like powers.   Now the details of the supersolid-to-superfluid transition as function of the non-ordering parameter, namely, the pressure $P$ and the flow $j_0$, will depend on the sign of $u_4$ in the present case. (It is to be noted that $u_6$ is taken to be positive as usual). But quite independently of these details, the effect of the zero-point energy and the pressure can be seen clearly from Eq. (9).  In order to approach the transition, the value of $r$ must decrease sufficiently. This is normally prevented here by the largeness of the zero-point term. It is, however, offset by the pressure term so as to bring about the transition. Most significantly now, once we are close enough to the transition point, a change in the flow $j_0$ can drive us across the transition as can be seen from the sign of the term quadratic in $j_0$ expression for $r$ in Eq. (9) that opposes the pressure term. Indeed, we can have then a flow-induced {\it melting} of the supersolid. This can be probed experimentally. The order parameter $\eta_g$ determined from the minimization of the free energy depends on the flow $j_0$, and enters the non-classical inertial mass as in Eq. (8). Inasmuch as for the potential flow in an annular geometry (1D flow with periodic boundary condition), the circulation $\oint({j_0}/{n_0})$ is quantized to $\nu h/m$, with $\nu$ an integer, this inertia will change in corresponding steps as indeed observed experimentally [1,2].

As for the nature of the transition, in the mean field we have $u_4 > 0$ give a second-order transition. For a first-order transition, we must have $u_4 < 0$ which is, of course, unphysical for a helium-like system with inter-atomic repulsive interaction. This pathology is due to the 1D model that has prevented the occurrence of the cubic term in the free energy in Eq. (8). 
 
Some general remarks are now in order. We have considered here only a potential (irrotational) flow and derived the translational non-classical inertia associated with the superflow ($j_0$) relative to the density modulation at rest in the laboratory frame. (A Galilean transformation can take us to a frame co-moving with the modulation relative to the laboratory frame). Experimentally, however, it is obviously convenient to have a bounded (confined) motion, which is readily realized in rotation $-$ hence the (non-classical) rotational inertia usually measured in experiments by confining the $^4$He-liquid (solidified under pressure) in an annulus which is then made to oscillate about its axis in a torsional mode. It is, however, to be noted that for an  annular thickness much smaller than the annular radius, the motion can be irrotational inasmuch as the annular region is not simply connected. Indeed, in the experiments of Kim and Chan [1-2], steps corresponding to the Onsager-Feynman quantization of circulation around annulus have been seen. Of course, we can have a situation where the motion has local circulation distributed in the form of vortices for a simply connected system $-$ $^4$He after all is a type II superfluid! In any case, fundamentally the translational inertia is well defined, calculable, and turns out to be non-classical as derived above. 

It is apt to recall here that a key point in the microscopic theory of superfluidity is the role of the condensate (macroscopic occupation of the zero-momentum single particle state), namely the hybridization of the single-particle excitation and the collective mode caused by the condensate [10]. In the present case, the complex modulated order parameter underlying the mass density wave is to be viewed as a macroscopically occupied single-particle state, and this too should subtend interesting hybridization effects. This calls for further study [11].

It may be apt to point out here that the problem of supersolids does raise certain general questions of interest about the partitioning of a given amount of angular momentum among the different possible modes of motion (degrees of freedom) so as to minimize the free energy, or just the energy at zero temperature. Thus, e.g., the angular momentum may be shared between the {\em orbital} and the {\em spin} motion for a system such as in $^3$He (assuming unpaired spins) giving induced spin polarization.  In the context of rotating superfluid $^4$He in a simply connected region, it may be taken up by the quantized vortices. Vortices and rotons have been invoked recently in the context of supersolids [12]. In a multiply connected (e.g., annular) region, the angular momentum may be taken up by the translational flow as in the present case. For a supersolid spinning about an axis, the angular momentum may be taken up by  delocalized point defects (the defectons where the number of lattice sites exceeds the number of $^4$He atoms [3]. These defectons may have local ring-like exchange motions. For a supersolid confined to a  thin rotating annulus, however, the quantum defects may carry the angular momentum by translating coherently around the annulus. We believe that atomistically this may be the case in the recent experiments [1,2]. The present single order-parameter based phenomenological theory, however, cannot address these atomistic details. 

In conclusion, we have constructed a supersolid model in terms of a complex order parameter underlying the real mass-density wave characteristic of a crystalline solid. This simple model gives a non-classical inertia smaller than the literal mass. It also gives the possibility of a flow-induced supersolid-to-superfluid transition. Finally, the supersolid as a non-uniform, modulated density $n(x)$ results from the fact that underlying the classical looking density modulation there is the quantum complex order parameter $\psi$ with $|\psi|^2 = n(x)$. Thus, a quantum supersolid is in a sense the square-root of a classical solid!

The author would like to thank P. Nozi\'{e}res for letting him have a copy of his very interesting preprint on this subject.

\end{document}